\documentclass[11pt]{article}

\usepackage[utf8]{inputenc}
\usepackage[T1]{fontenc}
\usepackage[a4paper,margin=1in]{geometry}
\usepackage[expansion=false]{microtype}
\usepackage{booktabs}
\usepackage{tabularx}
\usepackage{array}
\usepackage{ragged2e}
\usepackage{enumitem}
\usepackage{titlesec}
\usepackage{abstract}
\usepackage[hidelinks]{hyperref}
\usepackage{orcidlink}

\hypersetup{
  colorlinks=true,
  linkcolor=black,
  citecolor=black,
  urlcolor=blue,
  pdftitle={Bridging AI Risk Frameworks: Reconciling ISO/IEC 42001, the NIST AI RMF, and the EU AI Act into a Unified Governance Taxonomy},
  pdfauthor={Vinod Dhiman}
}

\newcolumntype{L}[1]{>{\RaggedRight\arraybackslash}p{#1}}

\titleformat{\section}{\normalfont\large\bfseries}{\thesection}{0.6em}{}
\titleformat{\subsection}{\normalfont\normalsize\bfseries}{\thesubsection}{0.6em}{}
\title{\vspace{-2.2em}\bfseries Bridging AI Risk Frameworks: Reconciling ISO/IEC 42001, the NIST AI Risk Management Framework, and the EU AI Act into a Unified Governance Taxonomy}

\author{%
  \textbf{Vinod Dhiman}\,\orcidlink{0009-0003-8358-9382}\\[0.3em]
  Technical Program Manager\\
  Arlington, VA, USA\\
  ORCID: 0009-0003-8358-9382\\
  Corresponding author: \href{mailto:vinod.dhiman@icloud.com}{vinod.dhiman@icloud.com}
}
\date{}

\begin{document}
\maketitle

\begin{abstract}
\noindent Artificial intelligence governance is consolidating around three structurally heterogeneous instruments: ISO/IEC 42001:2023, the first certifiable Artificial Intelligence Management System (AIMS) standard; the United States NIST AI Risk Management Framework (AI RMF 1.0), a voluntary, socio-technical risk model; and the European Union Artificial Intelligence Act (Regulation (EU) 2024/1689), a binding, risk-tiered law. Although these instruments share the goal of trustworthy AI, they differ fundamentally in legal status, governance subject, and conception of risk, so that the control-level crosswalks now common in practice are both incomplete and, in places, misleading. Drawing on document analysis of the official standards and frameworks and on comparative governance literature, this article reconciles the three instruments into a Unified AI Governance Taxonomy (UAGT) organized as five analytical layers -- normative purpose, governance subject, risk logic, control architecture, and evidence and assurance -- bound by a traceability spine and expressed through eight regulation-stable governance domains. The taxonomy is deliberately current, incorporating the May 2026 Digital Omnibus amendments to the AI Act and the 2024 NIST Generative AI Profile. A four-step implementation model and two worked high-risk examples -- an AI-enabled clinical decision-support system and an AI credit-scoring system -- show how organizations can operate a single control library and evidence base that supports ISO/IEC 42001 certification, NIST AI RMF adoption, and EU AI Act compliance without duplicative effort. We are explicit about where unification breaks down: the non-fungibility of legal conformity and voluntary certification, the mismatch between list-based and contextual risk ontologies, enforcement asymmetry, and the widening gap between all three instruments and general-purpose and agentic AI.

\vspace{0.6em}
\end{abstract}

\section{Introduction}

Organizations that design, develop, supply, and deploy artificial intelligence (AI) systems increasingly face overlapping yet structurally distinct governance expectations from standards bodies, regulators, and risk and compliance communities. The OECD AI Policy Observatory now tracks well over a thousand national and international AI policy initiatives, the overwhelming majority of which remain non-binding [40, 11]. Three instruments, however, have come to anchor the operational conversation for any organization operating across jurisdictions: ISO/IEC 42001:2023 [4]; the NIST AI Risk Management Framework [2]; and the EU Artificial Intelligence Act [1]. Each exerts influence well beyond its formal reach. The AI Act has an extraterritorial `Brussels effect' on any provider placing systems on the EU market [29]; the NIST framework, though voluntary, is increasingly treated by U.S. agencies and procurement processes as an expected baseline and has been incorporated into state law [20, 60]; and ISO/IEC 42001 has become the de facto certifiable expression of organizational AI governance maturity, jurisdiction-neutral and integrable with existing management systems [4, 31].

Although these artefacts share the goal of trustworthy AI, they differ in legal status, conceptual vocabulary, and operational emphasis. Without a harmonizing structure, organizations tend to build parallel, inconsistent governance tracks -- one for certification, one for internal risk management, and one for legal compliance -- producing duplicated controls, fragmented accountability, and avoidable cost [21, 22, 31, 61]. Practitioner crosswalks have proliferated in response, typically presenting clause-to-function-to-article tables that imply the three instruments are largely substitutable once aligned [30, 31, 32, 34, 37]. Such mappings are useful but tend to flatten three differences that are not artefacts of vocabulary and cannot be mapped away: the instruments differ in legal status (binding law versus voluntary guidance versus certifiable standard), in their governance subject (the AI system and its actor roles, the risk-management function and lifecycle, and the organization's management system, respectively), and in how they constitute risk itself (an ex ante, list-based categorization tied to intended purpose versus a contextual, measured, iterative assessment).

This article reconciles the three instruments into a layered model that supports both conceptual clarity and practical implementation at cloud scale. Rather than collapsing the frameworks into one another, the taxonomy identifies their shared governance primitives and maps their differences so that an organization can run a single control library and evidence base while respecting the distinct obligations each instrument imposes. The analysis is deliberately current, reflecting the state of the field in mid-2026 -- including the Digital Omnibus on AI agreed in May 2026, which materially rescheduled the AI Act's high-risk obligations [15, 16, 56], and the 2024 NIST Generative AI Profile together with NIST's ongoing revision of the base framework [3, 17].

The contributions are: (i) a conceptual methodology for integrating multi-framework AI governance based on document analysis and framework comparison; (ii) a comparative anatomy of the three instruments current to mid-2026; (iii) a Unified AI Governance Taxonomy with five analytical layers and eight regulation-stable governance domains, bound by a traceability spine; (iv) a four-step implementation model with two worked high-risk examples; and (v) an explicit account of the boundary conditions under which unification fails. Section 2 situates the work; Section 3 sets out the methodology; Section 4 presents the comparative anatomy; Section 5 separates convergence from divergence; Section 6 develops the taxonomy; Section 7 gives the implementation model and examples; Section 8 discusses implications; Section 9 sets out limitations and future work; and Section 10 concludes.

\section{Background and Related Work}

\subsection{AI governance as a systems problem}

AI governance can be understood as the structured discipline of inventorying AI systems, classifying their risk, applying proportionate controls, and monitoring them over time within a socio-technical context [27, 28]. Unlike traditional IT governance, AI governance must account for non-deterministic model behaviour, data quality and provenance, fairness and bias concerns, and complex human-in-the-loop workflows. The NIST AI RMF explicitly frames AI risk as socio-technical, emphasising that AI outcomes arise from interactions between technical components, human decision-makers, institutional contexts, and broader societal factors [2]. ISO/IEC 42001 embeds AI risk within a management-system cycle of policies, responsibilities, processes, and continual improvement, recognising that organisational structures and culture shape AI outcomes [4, 25]. The EU AI Act introduces a layered risk-classification model that distinguishes unacceptable, high, limited, and minimal AI uses, linking these categories to specific legal obligations [1, 51]. Table~\ref{tab:dimensions} summarizes the dimensions an enterprise AI governance programme must address across these views.

\begin{table}[htbp]
\centering
\caption{Key dimensions of AI governance in enterprise programmes.}
\label{tab:dimensions}
\small
\begin{tabularx}{\textwidth}{L{0.28\textwidth} X}
\toprule
\textbf{Dimension} & \textbf{Description} \\
\midrule
Technical robustness & Model validity, reliability, safety, security, and resilience across the AI lifecycle, including robustness to data drift and adversarial conditions [2, 7]. \\
Ethical \& societal impact & Fairness, non-discrimination, transparency, explainability, and broader social effects such as trust and legitimacy in AI-enabled decisions [27, 28]. \\
Legal \& regulatory compliance & Alignment with AI-specific regulation (e.g., the EU AI Act) and cross-cutting regimes such as data protection, consumer protection, and sectoral rules [1, 22]. \\
Organisational governance & Structures, roles, policies, and control processes that embed AI risk management into existing management systems (e.g., ISO 27001 and ISO/IEC 42001) [4, 39]. \\
Operational integration & Embedding governance into the SDLC, MLOps, and DevOps pipelines, including automation of controls, monitoring, and evidence collection [32, 34]. \\
\bottomrule
\end{tabularx}
\end{table}

\subsection{The lineage of risk-based regulation}

Risk-based regulation -- the targeting of regulatory effort at activities posing the greatest risk while lightening burdens on low-risk activity -- is a well-established governance technique that predates AI [23, 24]. Its application to AI is most visible in the EU AI Act, whose four tiers attach proportionate obligations to each band [1, 51]. Scholars observe that the Act's tiering is largely top-down and static: a system's tier depends primarily on whether its intended purpose appears on a pre-drawn list (notably Annex III) rather than on a measured probability and severity of harm in context [21, 22, 50]. This has prompted a substantial critical literature arguing that the categories may mis-estimate risk, struggle with general-purpose AI, and rely heavily on procedural conformity rather than substantive, ongoing assessment [47, 48, 52, 49, 53]. Against this legal lineage sits a managerial and engineering lineage: ISO 31000 established the general vocabulary of organizational risk management [10]; ISO/IEC 23894 adapted it to AI [7]; the NIST AI RMF translated risk management into four lifecycle functions [2]; and ISO/IEC 42001 enclosed these practices within the Plan-Do-Check-Act architecture shared with ISO/IEC 27001 and ISO 9001 [4, 25, 39]. Reconciling these two lineages -- one legal and rights-anchored, the other managerial and process-anchored -- is the central task of any unification effort.

\subsection{Fragmentation in practice and prior crosswalks}

Empirical and practitioner analyses indicate that organisations often treat ISO/IEC 42001, the NIST AI RMF, and the EU AI Act as distinct compliance tracks [31, 34]. Security and compliance teams may pursue ISO/IEC 42001 as an extension of ISO 27001; risk and engineering teams may adopt the NIST AI RMF for internal risk processes; legal teams may interpret AI Act obligations in isolation. The result is parallel documentation and fragmented artefacts -- separate AI inventories, risk registers, oversight procedures, and monitoring reports. A growing body of comparative work maps two or more instruments, often with automated gap analyzers and shared control registers [31, 32, 37, 38], and NIST itself has published crosswalks relating the AI RMF to international standards [18]. More academically, recent work has begun to formalize harmonization through reproducible crosswalks that decompose obligations into atomic, traceable unified controls organized by stable governance domains [32, 33]. Comparative governance studies widen the lens to include the OECD AI Principles, the Council of Europe Framework Convention, UNESCO's Recommendation, and the G7 Hiroshima Process as the normative substrate within which the three operational instruments sit [11, 12, 13, 14, 35, 45, 46, 54].

\subsection{Research gap}

Two gaps remain. First, most existing crosswalks operate at the level of control correspondence and therefore under-theorize the structural non-equivalences -- legal status, governance subject, and risk ontology -- that determine when a mapping is sound and when it silently fails. Second, the field lacks a synthesis that is simultaneously current (reflecting the 2024--2026 reshaping of all three instruments), layered (separating the normative, subject, risk, control, and evidentiary dimensions rather than collapsing them), and explicit about its own boundary conditions. This article addresses both.

A recent and closely related contribution, UGAF-ITS [33], consolidates the same three instruments into twelve unified controls across eight governance domains through a reproducible crosswalk methodology, but does so for the specific setting of distributed intelligent transportation systems. The present work is distinct in scope and construction. It is general and cloud-native rather than sector-specific; it separates a five-layer comparative lens (normative purpose, governance subject, risk logic, control architecture, and evidence and assurance) from the operative governance domains, rather than treating the domains as the whole model; it preserves all three risk verdicts rather than reconciling them into one; it is current to the May 2026 Digital Omnibus; and it is paired with an open, machine-readable, version-controlled implementation [63] so that the mappings can be maintained against changing deadlines and annexes. The eight regulation-stable domains used here are convergent with, but independently derived for, a broader enterprise and multi-jurisdiction setting.

\section{Methodology}

\subsection{Research approach}

This study adopts a design-oriented, conceptual research approach grounded in two complementary methods: document analysis of AI governance and regulatory sources, and architecture-driven synthesis of a governance model. The approach is consistent with design science in information systems, where artefacts such as taxonomies, models, and processes are constructed to address well-defined problem domains and evaluated in terms of utility and internal coherence rather than controlled experiment alone [32]. The objective is not to adjudicate the relative merits of the three instruments but to derive a unified governance taxonomy that reconciles them within a single, implementable structure for large, cloud-native organizations.

\subsection{Sources and selection}

Primary sources comprise three categories. First, official framework and standard texts: ISO/IEC 42001:2023 and its companion standards [4, 5, 6, 7, 9]; the NIST AI RMF 1.0 and the Generative AI Profile [2, 3]; and the consolidated AI Act with its 2026 amendments and Commission guidance [1, 15, 16]. Second, comparative and practitioner integration analyses describing how the three instruments combine into a single programme, particularly in cloud and SaaS environments [31, 32, 37, 38]. Third, peer-reviewed scholarship on risk-based regulation and responsible AI governance providing conceptual and methodological grounding [21, 22, 24, 25, 47, 53]. Sources were selected for authoritativeness, direct relevance to the three instruments, and applicability to enterprise-scale deployment.

\subsection{Synthesis procedure}

The synthesis proceeded in four stages. First, governance constructs -- principles, roles, lifecycle phases, risk concepts, control categories, and evidence requirements -- were extracted from the primary sources and coded inductively. Second, constructs were mapped across instruments to identify commonalities and divergences; for example, ISO/IEC 42001 leadership and context clauses were mapped to the NIST Govern function and to AI Act provider duties. Third, the mapped constructs were synthesised into five analytical layers and eight governance domains (Section 6), with each layer defined by how it appears in each instrument rather than by collapsing legal and voluntary artefacts into a single checklist. Fourth, the taxonomy was evaluated against documented multi-framework integration challenges -- fragmented systems, siloed ownership, regulatory complexity, and workflow integration [31, 34] -- and illustrated through two worked examples.

\subsection{Practitioner lens and limitations of the method}

Given the role of hyperscale cloud and SaaS platforms in AI deployment, the synthesis assumes a practitioner context of distributed, multi-cloud infrastructure with centralised security and compliance functions; existing management systems (e.g., ISO 27001, SOC 2) into which AI governance must integrate; and strong reliance on automation -- infrastructure as code, CI/CD, observability -- for operationalising controls and evidence [39, 34]. The resulting taxonomy is conceptual and has not yet been validated through longitudinal field studies or controlled experiments; the document analysis is necessarily selective; and the cloud-centric lens may require adaptation for other technology stacks or regulatory contexts. The methodology is nonetheless appropriate for producing a structured governance blueprint that can ground subsequent empirical evaluation and tooling.

\section{A Comparative Anatomy of the Three Frameworks}

\subsection{The EU AI Act: binding, risk-tiered, system-and-role oriented}

The Artificial Intelligence Act, Regulation (EU) 2024/1689, was adopted on 13 June 2024, published in the Official Journal on 12 July 2024, and entered into force on 1 August 2024 [1]. It is the world's first comprehensive, horizontal, binding AI law, adopting a risk-based architecture organized around an AI system's intended purpose [1, 51]. Practices posing unacceptable risk are prohibited under Article 5. High-risk systems, defined through Annex I (AI as a safety component of products already under EU harmonization law) and Annex III (stand-alone uses in domains such as biometrics, critical infrastructure, education, employment, essential services including creditworthiness assessment, law enforcement, and migration), carry the bulk of the substantive requirements: a risk-management system (Article 9), data and data governance (Article 10), technical documentation, record-keeping and logging, transparency to deployers, human oversight (Article 14), and accuracy, robustness, and cybersecurity (Article 15), followed by conformity assessment and post-market monitoring [1]. Limited-risk systems attract transparency obligations under Article 50, and minimal-risk systems are largely unregulated. Crucially, the Act regulates roles -- provider, deployer, importer, distributor -- and ties obligations to market placement and use, not to the maturity of an organization's internal management system.

The Act's phased application has become a moving target. Prohibitions and AI-literacy duties applied from 2 February 2025; obligations for general-purpose AI (GPAI) models and the governance architecture applied from 2 August 2025; and the bulk of the remaining provisions were scheduled for 2 August 2026 [1, 58]. The Digital Omnibus on AI -- proposed on 19 November 2025 [15] and agreed in trilogue on 7 May 2026 [16] -- then deferred the high-risk obligations: stand-alone Annex III systems now apply from 2 December 2027, and Annex I product-embedded systems from 2 August 2028, with national regulatory sandboxes pushed to 2 August 2027 [56, 57]. The same package introduced new Article 5 prohibitions targeting AI-generated non-consensual intimate imagery and child sexual abuse material, effective 2 December 2026, and adjusted transparency and watermarking grace periods [57, 55]. The substantive demands are thus stable in kind but volatile in timing -- a fact any durable taxonomy must absorb without rework.

\subsection{The NIST AI RMF: voluntary, function-oriented, lifecycle-based}

The NIST AI Risk Management Framework (AI RMF 1.0, NIST AI 100-1) was released on 26 January 2023 as voluntary guidance to help organizations incorporate trustworthiness into the design, development, use, and evaluation of AI [2]. It is organized around four iterative functions -- Govern, Map, Measure, and Manage -- and seven characteristics of trustworthy AI: valid and reliable; safe; secure and resilient; accountable and transparent; explainable and interpretable; privacy-enhanced; and fair, with harmful bias managed [2]. The framework is outcome-based rather than prescriptive, intended to remain applicable across a shifting technological landscape, and is accompanied by a Playbook and Roadmap [17]. On 26 July 2024 NIST released the Generative AI Profile (NIST AI 600-1), a cross-sectoral companion identifying twelve risk areas particular to or amplified by generative AI and proposing more than two hundred suggested actions mapped back to the four functions [3]. NIST has signalled continued evolution, revising AI RMF 1.0, while community efforts have proposed an agentic profile for autonomous, tool-using systems [17, 59]. Although voluntary, the framework is referenced by multiple U.S. agencies and was recognized in Colorado's original AI Act (SB 24-205), alongside ISO/IEC 42001, as a basis for an affirmative defense, although Colorado repealed and replaced that statute in 2026 [20]; the 2023 federal executive order that had promoted it was rescinded in January 2025, but the framework itself, as voluntary NIST guidance, remains in force and widely adopted [19].

\subsection{ISO/IEC 42001:2023: certifiable, organization-oriented, management-system based}

ISO/IEC 42001:2023, published in December 2023, is the first international standard specifying requirements for an artificial intelligence management system (AIMS) [4, 25]. It follows the ISO Harmonized Structure shared with ISO/IEC 27001 and ISO 9001, using Clauses 4 to 10 and the Plan-Do-Check-Act cycle [4, 39]. Its normative Annex A specifies thirty-eight controls grouped under nine control objectives, with Annex B providing implementation guidance, Annex C cataloguing AI-specific risk sources such as model drift and lack of explainability, and Annex D addressing sector applicability [23, 39]. The standard frames applicability through organizational roles -- AI provider, producer, customer, partner, subject, and relevant authority -- and requires a risk assessment (Clause 6.1) together with an AI system impact assessment where systems pose high potential impact to individuals, groups, or society, the latter elaborated in ISO/IEC 42005 [4, 8, 39]. Certification is voluntary, performed by accredited bodies, and typically valid for three years with annual surveillance audits [24, 26]. Its defining feature for comparative purposes is that it governs the organization's processes, not a specific system's market placement: it answers how an entity establishes, operates, and continually improves AI governance, leaving the substantive definition of unacceptable or high risk to be supplied by context, contract, or -- in the EU -- by law.

\subsection{Side-by-side comparison}

Table~\ref{tab:anatomy} summarizes the dimensions along which the three instruments differ. The differences in the first four rows -- legal status, governance subject, conception of risk, and the nature of the evidence produced -- are the structural facts any unification must respect rather than dissolve.

\begin{table}[htbp]
\centering
\caption{Comparative anatomy of the three anchor AI governance instruments (state of the field, mid-2026).}
\label{tab:anatomy}
\small
\begin{tabularx}{\textwidth}{L{0.16\textwidth} X X X}
\toprule
\textbf{Dimension} & \textbf{EU AI Act} & \textbf{NIST AI RMF} & \textbf{ISO/IEC 42001} \\
\midrule
Legal status & Binding regulation; directly applicable EU-wide & Voluntary guidance; no certification & Voluntary, certifiable international standard \\
Governance subject & The AI system and actor roles, by intended purpose & The risk-management function across the AI lifecycle & The organization's management system (AIMS) \\
Conception of risk & Ex ante, list-based tiers (Annex III); largely static & Contextual, measured, iterative (Map / Measure) & Organizational risk + AI system impact assessment via PDCA \\
Evidence produced & Conformity assessment, technical documentation & Documented function outcomes; self-attested alignment & Certification, Statement of Applicability, audit trail \\
Core structure & 113 articles, 13 annexes; chapters by risk tier & 4 functions: Govern, Map, Measure, Manage & Clauses 4--10; Annex A: 38 controls / 9 objectives \\
Treatment of GPAI & Dedicated GPAI / systemic-risk obligations (from Aug 2025) & Generative AI Profile (12 risk areas, 200+ actions) & Addressed via general controls; role-based scoping \\
Enforcement & Fines up to 7\% of global turnover; market surveillance & None directly; influence via agencies, procurement & Accredited certification; market and contractual pressure \\
Update cadence (current) & Digital Omnibus (May 2026) deferred high-risk deadlines & 1.0 under revision; new profiles (GenAI, critical infra) & Stable; expanding companion-standard family \\
\bottomrule
\end{tabularx}
\end{table}

\section{Convergence and Divergence}

\subsection{Genuine points of convergence}

Despite their different forms, the three instruments converge on a recognizable set of substantive expectations, which is precisely why a unified control library is feasible. Five convergences are robust. First, all three adopt a lifecycle orientation, treating governance as continuous rather than as a one-time gate [2, 4, 29]. Second, all three centre risk management as the organizing activity -- Article 9's risk-management system, the Map-Measure-Manage functions, and Clause 6.1's risk assessment [1, 2, 4]. Third, transparency and documentation are common load-bearing requirements, and transparency is widely identified as the most feasible site of transatlantic convergence [37, 35]. Fourth, human oversight and accountability appear throughout, from Article 14 to the Govern function to the standard's leadership clauses [1, 2, 4]. Fifth, all three increasingly attend to the value chain and third parties [3, 39, 33].

\subsection{Structural divergences that resist mapping}

These convergences sit atop divergences that control-level crosswalks tend to obscure. The most consequential is the non-fungibility of legal status. ISO/IEC 42001 certification demonstrates that an organization operates a conforming management system; it does not, by itself, establish conformity with the AI Act's substantive requirements for a given high-risk system. The bridge is the set of harmonized standards to be adopted under Article 40, which -- once published -- will allow conformity with a standard to confer a presumption of conformity with the law; until then, certification and legal conformity remain distinct artefacts [32, 36]. Treating a voluntary certificate as evidence of statutory compliance is a category error with liability consequences. A second divergence concerns the governance subject. The Act regulates systems and roles by intended purpose; the framework organizes around functions and the lifecycle; the standard governs the organization. A control satisfying the standard at the organizational level does not automatically discharge a system-specific obligation under Article 10, and a function-level NIST activity must be re-expressed as both an organizational control and a system-specific record to serve all three. The mapping is therefore many-to-many, not one-to-one [32, 34]. A third divergence is ontological: the Act's tiers are assigned ex ante by list membership tied to intended purpose, a top-down categorization criticized for insensitivity to measured, context-dependent harm [21, 22, 50, 53], whereas the framework treats risk as something to be mapped and measured in context and the standard treats it as an organizational quantity to be continually improved. A system can be `high-risk' as a matter of EU law while a contextual NIST measurement returns a comparatively modest residual risk -- or vice versa -- so a unified taxonomy must preserve, not average, the three verdicts. Finally, enforcement and scope diverge sharply, and enforcement asymmetry across member states and the self-assessment-heavy character of conformity have themselves been flagged as design weaknesses [1, 49].

\section{The Unified AI Governance Taxonomy (UAGT)}

\subsection{Design principles}

The taxonomy rests on four principles. (i) Layer separation: the normative, subject, risk, control, and evidentiary dimensions are kept distinct so that a change in one -- for example, a deferred AI Act deadline -- does not force re-architecting the others. (ii) Traceability: every unified control retains explicit back-links to its sources (an ISO/IEC 42001 clause or Annex A control, a NIST AI RMF subcategory, an AI Act article, and an underlying principle) and forward-links to evidence artefacts, following the reproducible-crosswalk logic emerging in the literature [32, 33] and realized in the accompanying open implementation [63]. (iii) Risk-verdict preservation: the taxonomy records all three risk determinations rather than collapsing them, so a legal high-risk classification is never silently overridden by a favourable contextual measurement. (iv) Regulation stability: the operative structure is a small set of governance domains that remain intact as deadlines, annexes, and profiles change.

Two structures work together. The five analytical layers are a comparative lens: each layer is a dimension along which all three instruments can be compared, avoiding the common error of equating a layer with a single framework. The eight governance domains are the operative content: the substantive control surface the instruments share. The traceability spine connects them -- a domain's controls link upward through the layers to a principle and outward to evidence.

\subsection{The five analytical layers}

Table~\ref{tab:layers} expresses the five layers across the three instruments. Read bottom-up, the layers guide design; read top-down, they yield an audit trail.

\begin{table}[htbp]
\centering
\caption{The five analytical layers of the UAGT mapped across the three instruments. Article references follow Regulation (EU) 2024/1689.}
\label{tab:layers}
\small
\begin{tabularx}{\textwidth}{L{0.14\textwidth} X X X}
\toprule
\textbf{Layer} & \textbf{ISO/IEC 42001} & \textbf{NIST AI RMF} & \textbf{EU AI Act} \\
\midrule
Normative purpose & Responsible AI management; ethical, transparent, accountable, safe use & Trustworthy AI: valid, safe, secure, resilient, accountable, explainable, privacy-enhanced, fair & Protection of health, safety, and fundamental rights; internal-market harmonisation \\
Governance subject & The organisation's AI management system (policies, processes, roles, records) & The AI system and its socio-technical context of use & AI systems as classified by risk; providers, deployers, importers, distributors \\
Risk logic & Organisational risk assessment plus AI impact assessment within PDCA & Dynamic, contextual, socio-technical risk across the lifecycle & Ex ante legal risk tiers (prohibited, high, limited, minimal) tied to intended purpose \\
Control architecture & Annex A: 38 controls / 9 objectives; documented procedures, audits, corrective action & Govern -- Map -- Measure -- Manage functions and outcomes & Prohibitions (Art 5); high-risk requirements (Arts 9--15); transparency (Art 50); conformity assessment; post-market monitoring \\
Evidence \& assurance & Certification, Statement of Applicability, audit trail & Risk registers, metrics, evaluations, governance-maturity artefacts & Technical documentation, logs, conformity declarations, serious-incident reports, fundamental-rights impact assessments \\
\bottomrule
\end{tabularx}
\end{table}

The traceability spine runs vertically through the layers. A single normative principle -- say, protection against discriminatory outcomes -- descends into an organizational fairness policy (governance subject), a method for measuring harmful bias (risk logic), a concrete obligation under Article 10 and the bias-detection provisions (control architecture), and a documented bias evaluation, model card, and impact-assessment record (evidence and assurance). Reading downward gives design guidance; reading upward gives the audit trail.

\subsection{Eight regulation-stable governance domains}

Layers describe the architecture; domains describe the operative content. We propose eight domains that, in our reading of the three instruments and the harmonization literature, remain stable as specific provisions evolve [32, 33]. Table~\ref{tab:domains} maps each domain to its principal expression in each framework.

\begin{table}[htbp]
\centering
\caption{Eight regulation-stable governance domains mapped to each instrument. Article references follow Regulation (EU) 2024/1689.}
\label{tab:domains}
\small
\begin{tabularx}{\textwidth}{L{0.17\textwidth} X X X}
\toprule
\textbf{Governance domain} & \textbf{EU AI Act} & \textbf{NIST AI RMF} & \textbf{ISO/IEC 42001} \\
\midrule
D1 Accountability \& organizational governance & Provider/deployer duties; quality management system (Art 17) & Govern function (roles, culture, policy) & Clauses 5 \& 4; Annex A governance controls \\
D2 Risk \& impact assessment & Risk-management system (Art 9) & Map \& Measure functions & Clause 6.1; AI impact assessment (ISO/IEC 42005) \\
D3 Data governance \& quality & Data and data governance (Art 10) & Map (context, data) \& Measure & Annex A data-for-AI controls \\
D4 Transparency, documentation \& records & Technical documentation, logging, deployer info (Arts 11--13, 50) & Accountable-and-transparent characteristic & Documented information; Statement of Applicability \\
D5 Human oversight \& autonomy & Human oversight (Art 14) & Human-AI configuration; Govern & Annex A oversight controls \\
D6 Robustness, accuracy \& security & Accuracy, robustness, cybersecurity (Art 15) & Safe; secure-and-resilient; valid-and-reliable & Annex A lifecycle and security controls \\
D7 Lifecycle monitoring \& post-market surveillance & Post-market monitoring; incident reporting (Arts 72, 73) & Manage function (ongoing) & Clause 9; Clause 10 (improvement) \\
D8 Value-chain, third-party \& GPAI governance & GPAI obligations; supply-chain duties & Value-chain \& component integration (GenAI Profile) & Annex A third-party / supplier controls \\
\bottomrule
\end{tabularx}
\end{table}

\section{Implementation Model}

\subsection{A four-step integration approach}

The taxonomy is operationalized through a four-step cycle aligned to ISO/IEC 42001's Plan-Do-Check-Act and the NIST Manage function. Step one, inventory and classification: build a central AI inventory aligned with the NIST Govern function and ISO/IEC 42001 context and scope, and classify each system under the AI Act's risk tiers, recording the legal verdict explicitly. Step two, reusable control sets: define control sets per risk tier that integrate ISO/IEC 42001 management-system requirements, NIST AI RMF outcomes, and AI Act obligations, parameterising jurisdiction-specific legal details while reusing a common core. Step three, evidence design: specify evidence artefacts -- policies, risk and impact assessments, test reports, logs, incident records -- that simultaneously support certification, internal governance, and regulatory compliance, placing each artefact at the layer where observability exists. Step four, operation and improvement: run the governance system as a continuous loop, updating controls as legal and technical contexts evolve. The output is a single cross-framework control register that can be presented, with appropriate framing, to a certification auditor, a counterparty assessing NIST alignment, and an EU market-surveillance authority.

\subsection{Aligning common challenges to the taxonomy}

Multi-framework integration raises recurring strategic, organizational, and technical challenges. Table~\ref{tab:challenges} maps the most common to the taxonomy layer at which they are best addressed and to an illustrative mitigation.

\begin{table}[htbp]
\centering
\caption{Common multi-framework integration challenges mapped to taxonomy layers and mitigations.}
\label{tab:challenges}
\small
\begin{tabularx}{\textwidth}{L{0.20\textwidth} L{0.20\textwidth} X}
\toprule
\textbf{Challenge} & \textbf{Taxonomy layer} & \textbf{Illustrative mitigation} \\
\midrule
Fragmented systems \& siloed ownership & Governance subject & Define a single AI inventory and a RACI spanning providers and deployers; map business units to one AIMS scope [31, 34]. \\
Conceptual misalignment & Risk logic & Adopt a shared risk vocabulary linking ISO risk cycles, NIST socio-technical risk, and AI Act tiers, preserving each verdict [32, 28]. \\
Regulatory complexity \& drift & Control architecture & Implement configurable control sets per risk tier and jurisdiction, reusing a core while parameterising legal specifics [56, 36]. \\
Workflow integration & Control architecture / Evidence & Embed governance checkpoints into CI/CD and MLOps; automate logging and documentation generation [32, 34]. \\
Manual processes \& scalability & Evidence \& assurance & Replace spreadsheet inventories with traceable control registers and automated evidence collection [31, 34]. \\
Skills \& interdisciplinary gaps & Normative purpose / Governance subject & Create cross-functional AI governance forums; clarify decision rights across legal, risk, and engineering [27, 28]. \\
\bottomrule
\end{tabularx}
\end{table}

\subsection{Worked example: an AI-enabled clinical decision-support system}

Consider a clinical decision-support system that triages imaging studies and prioritizes cases for radiologist review. It engages two layers of EU obligation simultaneously: as a medical device it falls under Annex I (product-embedded high-risk AI, now applying from 2 August 2028), and its triage function implicates Annex III health-related considerations [1, 56]. Under the UAGT, the organization first situates the system in the risk-logic and control-architecture layers by recording its legal status -- high-risk, subject to medical-device conformity assessment and to accuracy, robustness, human-oversight, and post-market-monitoring requirements -- then works the domains. In D2, the Article 9 risk-management exercise produces the NIST Map/Measure outputs and the ISO/IEC 42005 impact assessment as one artefact linked three ways. In D3, data-governance evidence serves Article 10, the NIST data-context activities, and the standard's data controls. In D5, human oversight is documented as a system-specific design feature -- the radiologist remains the decision-maker -- discharging Article 14 and the human-AI configuration concern. In D6, clinical validation, robustness testing, and cybersecurity hardening form one evidence set mapped to Article 15, the safe and secure-and-resilient characteristics, and the relevant Annex A controls. In D7, post-market monitoring and serious-incident reporting satisfy Articles 72 and 73, the Manage function, and Clauses 9 and 10. The organization preserves all three risk verdicts: the system is legally high-risk regardless of whether a contextual NIST measurement of residual risk is low.

\subsection{Worked example: an AI credit-scoring system}

Consider next a financial institution deploying an AI credit-scoring system, explicitly listed among high-risk uses in Annex III. The four-step model applies directly. At inventory and classification, the system is registered and tagged high-risk under the Act. At reusable control sets, ISO/IEC 42001 supplies the organizational backbone -- policies on model development, data governance, human oversight, and incident reporting -- while the NIST functions structure risk identification, mapping, measurement, and management, and the AI Act fixes the binding obligations on documentation, oversight, logging, and incident reporting. At evidence design, a single set of artefacts (model documentation, bias and performance evaluations, decision logs, and a fundamental-rights impact assessment) serves certification, internal governance, and conformity simultaneously. At operation and improvement, drift monitoring and periodic review feed the PDCA loop and the Manage function. As in the clinical case, the institution records the legal high-risk verdict alongside its contextual and organizational risk assessments rather than letting a favourable internal measurement displace the statutory obligation [3, 32].

\section{Discussion}

\subsection{Implications for practice}

The central practical pay-off is the single cross-framework control register: one control library, one evidence base, and one set of processes, each item traceable to every instrument it serves [31, 34]. This reduces the recurring administrative cost of parallel programmes and the risk that a control satisfying one framework is assumed to satisfy another. The layered design also clarifies sequencing: organizations can begin with the ISO/IEC 42001 management-system container, adopt the NIST functions as their risk method, and map the AI Act's obligations as the binding floor -- a sequence consistent with guidance that treats the standard as the vehicle through which AI Act requirements are operationalized [31, 36]. Because controls are atomic and traceable, automated gap analysis and evidence collection become tractable [34, 37].

\subsection{Implications for policy}

At the policy level, the analysis underscores the decisive role of Article 40 harmonized standards as the formal bridge between voluntary conformity and legal presumption; the slow arrival of these standards was itself a stated reason for the Digital Omnibus deferrals, which linked high-risk applicability to the availability of supporting standards and tools [15, 56, 57, 36]. It also illustrates the layered international architecture within which the three instruments sit: the OECD Principles and the Council of Europe Framework Convention supply the normative layer, the Convention adding binding human-rights obligations that the operational instruments presuppose but do not themselves create [11, 12, 39, 44, 46]. The broader trajectory is convergence through interaction rather than consolidation, as the Convention and the Act both align with the OECD definition of an AI system [39, 35]. Recent policy critique likewise argues for recalibrating, rather than abandoning, the EU's approach [62, 42].

\subsection{The Brussels effect and interoperability}

Because the AI Act binds any provider placing systems on the EU market, its requirements propagate globally through firms that standardize on the strictest applicable regime -- the Brussels effect [29]. The UAGT operationalizes a constructive version of this dynamic: an organization satisfies the binding floor where it applies while using the jurisdiction-neutral layers as a portable substrate that travels across markets, consistent with the OECD's framing of interoperability as the antidote to fragmentation [40, 45] and with calls to coordinate the EU, U.S., OECD, G7, and Council of Europe efforts [35, 41, 43].

\section{Limitations and Future Work}

The taxonomy has boundary conditions that should be stated plainly. First, unification is structural, not legal: the UAGT helps an organization reuse effort and evidence, but it cannot convert ISO/IEC 42001 certification or NIST alignment into AI Act conformity; only Article 40 harmonized standards, once adopted, will confer a presumption of conformity [32, 36]. Second, the risk-ontology mismatch is irreducible: list-based legal tiers and contextual measured risk can disagree, and the taxonomy deliberately preserves rather than reconciles those verdicts [21, 50, 53]. Third, the analysis is time-stamped: the Digital Omnibus reset key deadlines in 2026, NIST is revising its base framework, and ISO's companion-standard family is still expanding, so the control-architecture and evidence layers will require periodic re-mapping even though the domains remain stable [15, 57, 17]. Fourth, all three instruments are being outrun by the frontier -- general-purpose and increasingly agentic systems raise governance questions that the AI Act's GPAI provisions, the NIST Generative AI Profile, and the standard's general controls address only partially [3, 59].

Three lines of future work follow. The first is empirical validation: applying the UAGT to a portfolio of real systems and measuring the reduction in duplicated controls and evidence, and the rate of detected gaps, against a siloed baseline. The second is formalization: the atomic-obligation crosswalk is expressed in a machine-readable, schema-validated, version-controlled implementation [63] so that mappings can be maintained against changing deadlines and annexes; future work will extend its coverage and validation. The third is extension: incorporating sectoral regimes (medical devices, financial services, critical infrastructure) and emerging agentic-AI guidance as additional, traceable sources within the existing layer-and-domain structure rather than as separate frameworks [33].

\section{Conclusion}

ISO/IEC 42001, the NIST AI RMF, and the EU AI Act are not three answers to the same question; they answer three different questions -- how an organization should govern AI, how it should manage AI risk, and what it must legally do -- that share a large substantive surface. The temptation to treat them as interchangeable, encouraged by the abundance of control-level crosswalks, obscures structural non-equivalences in legal status, governance subject, and risk ontology that determine when alignment is sound. The Unified AI Governance Taxonomy proposed here lets organizations capture the genuine overlap -- a single, traceable control library and evidence base organized into five analytical layers and eight regulation-stable domains -- while preserving the distinct obligations and risk verdicts each instrument imposes, and operationalizes it through a four-step model demonstrated on two high-risk use cases. In a landscape that is converging through interaction rather than consolidation, and that is reshaped almost yearly by amendments such as the 2026 Digital Omnibus, the value of such a taxonomy lies less in declaring the frameworks unified than in making their differences legible and their commonalities reusable.

\section*{Declarations}

\textbf{Funding.} The author received no specific funding for this work.

\textbf{Competing interests.} The author declares no competing interests. The views expressed are solely those of the author and do not necessarily reflect the positions of any current or past employer, client, or affiliated institution. No funding, sponsorship, or in-kind support was received from any organisation with a direct stake in ISO/IEC 42001, the NIST AI RMF, or EU AI Act implementations.

\textbf{Data availability.} Not applicable; this is a conceptual and comparative study drawing on publicly available instruments and literature. The accompanying machine-readable crosswalk is openly available [63].

\textbf{Author contributions.} The sole author conceived, researched, and wrote the manuscript.

\section*{References}
\small
\begin{enumerate}[label={[\arabic*]}, leftmargin=2.4em, itemsep=2pt]
\item European Parliament and Council of the European Union. Regulation (EU) 2024/1689 laying down harmonised rules on artificial intelligence (Artificial Intelligence Act). \textit{Official Journal of the European Union}, L series, 12 July 2024.
\item National Institute of Standards and Technology. Artificial Intelligence Risk Management Framework (AI RMF 1.0), NIST AI 100-1. Gaithersburg, MD: U.S. Department of Commerce, 26 January 2023. \url{https://doi.org/10.6028/NIST.AI.100-1}
\item Autio, C., Schwartz, R., Dunietz, J., Jain, S., Stanley, M., Tabassi, E., Hall, P., Roberts, K. Artificial Intelligence Risk Management Framework: Generative Artificial Intelligence Profile, NIST AI 600-1. NIST, 26 July 2024. \url{https://doi.org/10.6028/NIST.AI.600-1}
\item ISO/IEC. ISO/IEC 42001:2023 -- Information technology -- Artificial intelligence -- Management system. Geneva: ISO, December 2023.
\item ISO/IEC. ISO/IEC 22989:2022 -- Information technology -- Artificial intelligence -- Artificial intelligence concepts and terminology. Geneva: ISO, 2022.
\item ISO/IEC. ISO/IEC 23053:2022 -- Framework for Artificial Intelligence (AI) Systems Using Machine Learning (ML). Geneva: ISO, 2022.
\item ISO/IEC. ISO/IEC 23894:2023 -- Information technology -- Artificial intelligence -- Guidance on risk management. Geneva: ISO, 2023.
\item ISO/IEC. ISO/IEC 42005:2025 -- Information technology -- Artificial intelligence -- AI system impact assessment. Geneva: ISO, 2025.
\item ISO/IEC. ISO/IEC 38507:2022 -- Governance of IT -- Governance implications of the use of artificial intelligence by organizations. Geneva: ISO, 2022.
\item International Organization for Standardization. ISO 31000:2018 -- Risk management -- Guidelines. Geneva: ISO, 2018.
\item Organisation for Economic Co-operation and Development. Recommendation of the Council on Artificial Intelligence (OECD AI Principles), OECD/LEGAL/0449. Adopted 2019, updated May 2024.
\item Council of Europe. Framework Convention on Artificial Intelligence and Human Rights, Democracy and the Rule of Law (CETS No. 225). Adopted May 2024; opened for signature 5 September 2024.
\item UNESCO. Recommendation on the Ethics of Artificial Intelligence. Paris: UNESCO, November 2021.
\item G7. Hiroshima Process International Guiding Principles for Organizations Developing Advanced AI Systems and International Code of Conduct. 2023.
\item European Commission. Proposal for a Regulation amending Regulations (EU) 2024/1689 and (EU) 2018/1139 as regards the simplification of the implementation of harmonised rules on artificial intelligence (Digital Omnibus on AI). COM(2025) 836 final, 2025/0359(COD). Brussels, 19 November 2025.
\item Council of the European Union. Artificial intelligence: Council and Parliament agree to simplify and streamline rules (press release). 7 May 2026.
\item National Institute of Standards and Technology. AI RMF Playbook and Roadmap. NIST AI Resource Center, 2023--2024.
\item National Institute of Standards and Technology. AI RMF Crosswalk Documents (mappings of the AI RMF to ISO/IEC 42001, the EU AI Act, the OECD AI Principles, and other frameworks). NIST AI Resource Center (AIRC). \url{https://airc.nist.gov/airmf-resources/crosswalks/}
\item Executive Office of the President. Executive Order 14110: Safe, Secure, and Trustworthy Development and Use of Artificial Intelligence. Federal Register, 30 October 2023. Rescinded 20 January 2025 by Executive Order 14148 and superseded by Executive Order 14179 (23 January 2025).
\item Colorado General Assembly. Senate Bill 24-205, Consumer Protections for Artificial Intelligence (Colorado AI Act), 2024. Effective date delayed from 1 February 2026 to 30 June 2026 by SB 25B-004 (28 August 2025), then repealed and replaced by SB 26-189 (signed 14 May 2026; effective 1 January 2027), which removed the NIST/ISO affirmative-defense provision.
\item Mahler, T. Between Risk Management and Proportionality: The Risk-Based Approach in the EU's Artificial Intelligence Act Proposal. In: Colonna, L., Greenstein, R. (eds) Nordic Yearbook of Law and Informatics 2020--2021. Stockholm: Swedish Law and Informatics Research Institute, 2022, pp. 247 ff. \url{https://doi.org/10.53292/208f5901.38a67238}
\item Chamberlain, J. The Risk-Based Approach of the European Union's Proposed Artificial Intelligence Regulation: Some Comments from a Tort Law Perspective. \textit{European Journal of Risk Regulation}, 2023. \url{https://doi.org/10.1017/err.2022.38}
\item Black, J. Risk-Based Regulation: Choices, Practices and Lessons Being Learnt. In: OECD (ed.) Risk and Regulatory Policy: Improving the Governance of Risk. Paris: OECD Publishing, 2010, p. 187. \url{https://doi.org/10.1787/9789264082939-en}
\item Baldwin, R., Black, J. Driving Priorities in Risk-Based Regulation: What's the Problem? \textit{Journal of Law and Society}, 2016, 43(4): 565--595. \url{https://doi.org/10.1111/jols.12003}
\item Smuha, N.A. Beyond a Human Rights-Based Approach to AI Governance: Promise, Pitfalls, Plea. \textit{Philosophy \& Technology}, 2021, 34: 91--104. \url{https://doi.org/10.1007/s13347-020-00403-w}
\item Veale, M., Zuiderveen Borgesius, F. Demystifying the Draft EU Artificial Intelligence Act. \textit{Computer Law Review International}, 2021, 22(4): 97--112. \url{https://doi.org/10.9785/cri-2021-220402}
\item Yeung, K., Howes, A., Pogrebna, G. AI Governance by Human Rights-Centred Design, Deliberation and Oversight: An End to Ethics Washing. In: Dubber, M., Pasquale, F. (eds) The Oxford Handbook of AI Ethics. Oxford: Oxford University Press, 2019.
\item Floridi, L., Cowls, J., Beltrametti, M., et al. AI4People -- An Ethical Framework for a Good AI Society. \textit{Minds and Machines}, 2018, 28: 689--707. \url{https://doi.org/10.1007/s11023-018-9482-5}
\item Bradford, A. The Brussels Effect: How the European Union Rules the World. New York: Oxford University Press, 2020.
\item Sprinto. Your Guide to ISO 42001 Controls for AI Governance. 2026. (Practitioner source.)
\item Truvo Cyber. ISO 42001 and the EU AI Act: What Actually Maps and What Doesn't. 2026. (Practitioner source.)
\item Sharma, A.G. ISO 42001 \& NIST AI RMF to EU AI Act: Framework Mapping Guide (clause-by-clause crosswalk). EU AI Compass, 18 March 2026. (Practitioner source.)
\item Butt, T.A., et al. UGAF-ITS: A Standards Harmonization Framework and Validation Tool for Multi-Framework AI Governance in Distributed Intelligent Transportation Systems. arXiv preprint, 2026. arXiv:2604.22789.
\item TechAhead. NIST AI RMF: A Practical Implementation Guide (crosswalk of NIST subcategories to ISO 42001 and EU AI Act). 2026. (Practitioner source.)
\item Roberts, H., Hine, E., Taddeo, M., Floridi, L. Global AI governance: barriers and pathways forward. \textit{International Affairs}, 2024, 100(3): 1275--1286. \url{https://doi.org/10.1093/ia/iiae073}
\item Training Camp / IAPP. How AIGP Maps to the EU AI Act, NIST AI RMF, and ISO 42001. 2026. (Practitioner source.)
\item EC-Council. EU AI Act, NIST AI RMF and ISO/IEC 42001: A Plain English Comparison. 2026. (Practitioner source.)
\item FairNow. Integrating the NIST AI RMF and ISO 42001: A Practical Guide. 2025. (Practitioner source.)
\item Council of Europe / Regulations.AI. Commentary on the Framework Convention on AI (CETS No. 225) and its alignment with the OECD definition of an AI system. 2026.
\item Organisation for Economic Co-operation and Development. OECD AI Policy Observatory: national and international AI policy database. 2025.
\item ENSURED. Global AI Regulation at a Time of Transformation: The Council of Europe's Framework Convention on Artificial Intelligence (Research Report). 2025.
\item ENSURED. Anchoring Global AI Governance: How the EU Can Leverage the Council of Europe's Framework Convention on AI (Policy Brief). 2025.
\item Vardiashvili, T. The Framework Convention on AI: A New Era in Global Tech Governance. European Studies Review, 29 January 2025.
\item European Parliament. Recommendation on the draft Council decision on the conclusion, on behalf of the European Union, of the Council of Europe Framework Convention on Artificial Intelligence and Human Rights, Democracy and the Rule of Law (A10-0007/2026). Adopted 11 March 2026.
\item Dogan, O.E. AI Regulation Across the Atlantic: EU AI Act vs. U.S. AI Governance. Global Policy Institute (GPI), 2025.
\item University of Illinois Law Review. The First Global AI Treaty (online essay). December 2024.
\item Fraser, H.L., Bello y Villarino, J.-M. Acceptable Risks in Europe's Proposed AI Act: Reasonableness and Other Principles for Deciding How Much Risk Management Is Enough. \textit{European Journal of Risk Regulation}, 2023. \url{https://doi.org/10.1017/err.2023.57}
\item Novelli, C., Casolari, F., Rotolo, A., Taddeo, M., Floridi, L. Taking AI risks seriously: a new assessment model for the AI Act. \textit{AI \& Society}, 2024, 39(5): 2493--2497. \url{https://doi.org/10.1007/s00146-023-01723-z}
\item Smuha, N.A., Yeung, K. The European Union's AI Act: Beyond Motherhood and Apple Pie? In: Smuha, N.A. (ed) The Cambridge Handbook on the Law, Ethics and Policy of Artificial Intelligence. Cambridge: Cambridge University Press, 2025, p. 33.
\item Ebers, M. Truly Risk-Based Regulation of Artificial Intelligence -- How to Implement the EU's AI Act. \textit{European Journal of Risk Regulation}, 2025, 16(2): 684--703. \url{https://doi.org/10.1017/err.2024.78}
\item EU AI Act knowledge base. Key Issue 3: Risk-Based Approach; Article 6 classification rules. artificialintelligenceact.eu, 2024--2026.
\item Schuett, J. Risk Management in the Artificial Intelligence Act. \textit{European Journal of Risk Regulation}, 2023. \url{https://doi.org/10.1017/err.2023.1}
\item Novelli, C., Casolari, F., Rotolo, A., Taddeo, M., Floridi, L. AI Risk Assessment: A Scenario-Based, Proportional Methodology for the AI Act. \textit{Digital Society}, 2024. \url{https://doi.org/10.1007/s44206-024-00095-1}
\item Du, J. Toward Responsible and Beneficial AI: Comparing Regulatory and Guidance-Based Approaches -- A Comprehensive Comparative Analysis of Artificial Intelligence Governance Frameworks across the European Union, United States, China, and IEEE. arXiv preprint, 2025. arXiv:2508.00868.
\item Stibbe. AI Act reloaded? What the latest AI Act changes mean in practice. 2026.
\item Gibson Dunn. EU AI Act Omnibus Agreement -- Postponed High-Risk Deadlines and Other Key Changes. 2026.
\item Covington \& Burling (Inside Privacy / Inside Global Tech). EU AI Act Update: Timeline Relief, Targeted Simplification, and New Prohibitions. May 2026.
\item Future of Life Institute. EU Artificial Intelligence Act -- Implementation Timeline. artificialintelligenceact.eu, 2026.
\item Cloud Security Alliance. NIST AI Risk Management Framework: Agentic Profile (v1, whitepaper). 2026.
\item IAPP. AI Governance Profession Report 2025. 2025.
\item EY. Responsible AI Pulse Survey 2025 (C-suite leaders across 21 countries). 2025.
\item Bruegel. The right balance: how to fix European Union artificial intelligence regulation (policy brief). 2026.
\item Dhiman, V. UAGT -- Unified AI Governance Taxonomy: an open, machine-readable crosswalk across ISO/IEC 42001, the NIST AI RMF, and the EU AI Act. Zenodo, 2026. \url{https://doi.org/10.5281/zenodo.20823079}
\end{enumerate}

\end{document}